\documentstyle[multicol,aps,pre,epsf,amstex]{revtex}
\begin{document}
\title{Freezing in Ising Ferromagnets}
\author{V.~Spirin, P.~L.~Krapivsky, and S.~Redner}

\address{Center for BioDynamics, Center for Polymer Studies,
and Department of Physics, Boston University, Boston, MA, 02215}
\maketitle
\begin{abstract}
  
  We investigate the final state of zero-temperature Ising ferromagnets which
  are endowed with single-spin flip Glauber dynamics.  Surprisingly, the
  ground state is generally {\em not\/} reached for zero initial
  magnetization.  In two dimensions, the system either reaches a frozen
  stripe state with probability $\approx 1/3$ or the ground state with
  probability $\approx 2/3$.  In greater than two dimensions, the probability
  to reach the ground state or a frozen state rapidly vanishes as the system
  size increases and the system wanders forever in an iso-energy set of
  metastable states.  An external magnetic field changes the situation
  drastically -- in two dimensions the favorable ground state is always
  reached, while in three dimensions the field must exceed a threshold value
  to reach the ground state.  For small but non-zero temperature, relaxation
  to the final state first proceeds by the formation of very long-lived
  metastable states, similar to the zero-temperature case, before equilibrium
  is reached.

\end{abstract}
\begin{multicols}{2}

\section{Introduction}
\subsection{Background}

Despite extensive study\cite{glauber,rev}, several fundamental questions
about the kinetic Ising model with Glauber dynamics still remain unanswered
\cite{ns}.  In this paper, we investigate the following basic issue: What is
the final state of a finite Ising-Glauber spin system when it is suddenly
quenched from infinite temperature to zero temperature\cite{skr}?  An
infinite system undergoes coarsening, that is, the spins organize into a
coarsening domain mosaic of up and down spins, with the characteristic domain
length scale growing as $t^{1/2}$.  For a finite system, this coarsening is
interrupted when the typical domain size reaches the system size $L$.  Rather
unexpectedly, this interruption can cause the system to get stuck in an
infinitely long-lived metastable state.  In two dimensions, it appears that,
as $L\to\infty$, this probability of getting stuck is approximately $1/3$,
while for $d\geq 3$ the ground state is never reached.

To provide context for the question of the ultimate fate of the Ising-Glauber
system, consider the limiting cases of $d=1$ \cite{skr,liggett} and
$d=\infty$.  For a linear chain of length $L$ which initially contains $pL$
up spins and $(1-p)L$ down spins, the final state of the system follows from
two facts.  First, the average magnetization is conserved under Glauber
dynamics \cite{glauber,liggett} and second, the only possible final states
are all spins up or all spins down; there are no metastable states in one
dimension.  To achieve a final magnetization which equals the initial
magnetization, a fraction $p$ of all realizations must end with all spins up
and the complementary fraction must end with all spins down.  Thus the
probability $P(p)$ to ultimately have all spins up as a function of the
initial concentration of up spins $p$ is simply $P(p)=p$.  A very different
result holds in the mean-field limit.  A simple realization of this limit is
the complete graph, in which each spin interacts with every other spin in the
system.  Here, a non-zero magnetization makes it energetically favorable for
any minority spin to flip, so that the majority phase quickly fills the
system for all $p\ne 1/2$.  Therefore on the complete graph, $P(p)$ is simply
the step function $P(p)=\theta(p-1/2)$.

We argue that in two and higher dimensions the probability that one phase
eventually wins also converges to a step function in $p-1/2$ but with strange
anomalies when $p=1/2$.  As mentioned above, the system has a non-zero
probability of getting stuck in a metastable state which consists of two or
more straight stripes in two dimensions.  In greater than two dimensions, the
probability that the ground state is reached quickly vanishes with system
size.  Intriguingly, the final state is not static, but rather consists of
stochastic ``blinkers''.  These are localized sets of spins which can flip
{\it ad infinitum} without any energy cost.  Thus the system wanders forever
on a connected set of equal-energy states defined by these blinkers.  In the
categorization proposed by Newman and Stein\cite{ns}, the Ising system in
high spatial dimension is of type ${\cal M}$ (``mixed'') in that a fraction of
the spins flip a finite number of times, while the complementary fraction
flip infinitely often.

We also study the fate of the Ising system in the presence of an external
magnetic field.  The imposition of a field leads to bootstrap percolation
phenomena\cite{bootstrap,perc}.  In two dimensions, an infinitesimal field is
sufficient to drive the system to the energetically favorable ground state;
this is the analog of the percolation threshold being equal to 1 in $n=3$
bootstrap percolation on the square lattice.  In three dimensions, the
energetically favorable ground state is reached only if the field exceeds a
threshold value, while for weaker fields, a transition between phase
coexistence and field alignment occurs as a function of the initial
concentration of field-aligned spins.  This transition again can be described
in terms of bootstrap percolation.

Finally, we examine the relaxation at non-zero temperatures.  While the
system eventually reaches equilibrium, the aforementioned metastable states
continue to play a significant role in the relaxation.  For example in $d=2$
and temperatures up to approximately $0.2T_c$, there is a large time range
for which the relaxation is close to that of the zero-temperature system.  If
a metastable stripe configuration happens to form, a time of the order of
$L^3\,e^{4J/kT}$, where $J$ is the interaction strength between spins, must
elapse before the system can escape this metastable state and reach
equilibrium.  We expect that analogous metastable states will influence the
long-time relaxation even more strongly in higher dimensions.

\subsection{The Model}

We study the homogeneous ferromagnetic Ising model with Hamiltonian ${\cal
  H}=-J\sum_{\langle ij\rangle}\sigma_i\sigma_j$, where $\sigma_i=\pm 1$ and
the sum is over all nearest-neighbor pairs of sites $\langle i j\rangle$.
The spins are initially uncorrelated, corresponding to an initial temperature
$T=\infty$.  The $T=\infty$ limit implies that the fraction of up and down
spins are equal.  Because we are interested in subtle effects associated with
zero initial magnetization, we prepare our systems with fixed magnetization,
rather than fixed probability for the orientation of each spin.  Thus every
initial configuration has $L^d/2$ up spins and $L^d/2$ down spins.  We also
consider the straightforward generalization to non-zero magnetization by
having $p\times L^d$ up spins and $(1-p)\times L^d$ down spins, where $p$ is
the initial fraction of up spins.

The spins evolve by zero-temperature Glauber dynamics\cite{glauber}, {\it i.\ 
  e.}, we view the system as being suddenly quenched from $T=\infty$ to
$T=0$.  (In Sect.~IV, we will also consider the $T>0$ case.)~ For each
initial spin configuration, one realization of the dynamics was performed
until the final state was reached.  We focus on $d$-dimensional hypercubic
lattices with linear size $L$ and periodic boundary conditions.  We choose
these lattices to avoid the pathologies associated with odd-coordinated
lattices, where stable convex islands of minority spins readily form and the
system always freezes into a disordered state\cite{ns,skr}.  Most of our
results continue to hold for free boundary conditions and on arbitrary
even-coordinated lattices.

Glauber dynamics at zero temperature is implemented by picking a spin at
random and computing the energy change $\Delta E$ if this spin were to flip.
If $\Delta E$ is greater than, equal to, or less than zero, 
the spin flip is accepted with probability $1$,
$1/2$ and $0$, respectively.  After each such event, the time is incremented
by $1/L^d$.  At long times this procedure is prohibitively slow because there
are relatively few flippable spins -- those with $\Delta E\leq 0$.  Thus we
track and randomly pick only flippable spins, and then update the time by
$1/(\hbox{number of flippable spins})$.  This implementation guarantees that
in one time unit, each spin attempts one flip on average.

The rest of this paper is organized as follows.  In Sect.~II we describe
geometric properties of the final state in various spatial dimensions
including the distribution of magnetization and energy in the final state, as
well as the influence of an external magnetic field in determining the final
state.  The number of metastable states as a function of the
spatial dimension is estimated in Sect.~III.  In Sect.~IV, we discuss the
finite-temperature evolution in two and three dimensions.  We conclude in
Sect.~V with a summary and some open questions.

\section{Final State Geometry}

We first address the following basic question: what is the probability $P(p)$
that the Ising system with $p\times L^d$ up spins in the initial state which
is then suddenly quenched to $T=0$ will ultimately have all spins up?  One
could imagine three possible outcomes:
\begin{itemize}
\item[1.]
$P(p)$ is a monotonically increasing function of $p$.
\item[2.] $P(p)=0$ for sufficiently small initial concentration of up spins
  $0\leq p\leq p_c$; $P(p)$ increases monotonically for $p_c<p<1-p_c$;
  $P(p)=1$ for $p\geq 1-p_c$.
\item[3.]
$P(p)$ is the step function $P(p)=\theta(p-1/2)$.
\end{itemize}
In one dimension $P(p)=p$, {\it i.\ e.}, the first scenario applies.  In this
section we shall argue that in higher dimensions the third scenario is
realized.

If the third scenario applies, then additional considerations are needed to
determine $P(1/2)$.  The behavior at $p=1/2$ is of paramount interest because
in a genuine $T=\infty$ initial state the concentrations of up and down spins
are equal (when initial temperature exceeds $T_c$, these concentrations are
also equal as long as there is no external magnetic field).  A general belief
is that in the thermodynamic limit the system always reaches one of the
ground states.  The symmetry between up and down phases then implies
$P(1/2)=1/2$. This assertion turns out to be {\em wrong} for all $d\geq 2$;
the two-dimensional system does not always reach the ground state while in
higher dimensions the system never reaches the ground state as $L\to\infty$.

\begin{figure}
  \narrowtext \epsfxsize=1.2in\hskip 0.9in\epsfbox{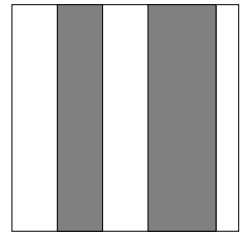} \vskip
  0.25in
\caption{Four-stripe metastable state in two dimensions.
\label{simple}}
\end{figure}

The crucial difference between one and higher dimensions is that no
metastable states exist in one dimension while there are numerous metastable
states in higher dimensions.  The existence of such
states\cite{other,lif,saf} is easy to visualize in two dimensions, where any
stripe of width $\geq 2$ which traverses the entire system is obviously
stable at zero temperature (Fig.~\ref{simple}).  However, it is not a priori
clear what is the basin of attraction of these metastable states and the
relative size of this basin compared to the basin of attraction to the ground
states.  We now turn to numerical results which indicate that metastable
states profoundly affect the fate of arbitrarily large higher-dimensional
zero-temperature Ising systems.

\subsection{Two Dimensions}

\subsubsection{Stripe state in zero magnetic field}

\begin{figure}
  \narrowtext \epsfxsize=2.8in\hskip 0.0in\epsfbox{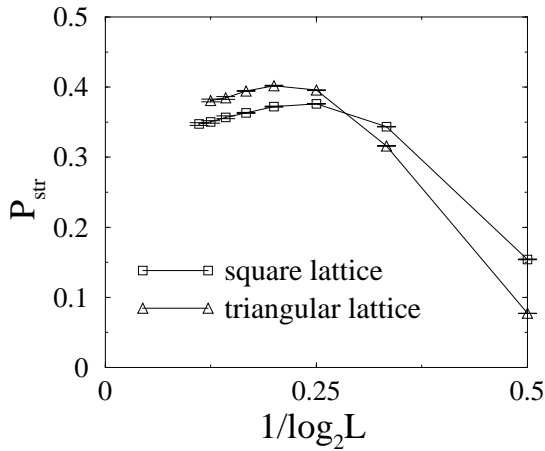} \vskip
  0.25in
\caption{Probability that an $L\times L$ system ($\square$ square lattice,
  $\Delta$ triangular lattice) eventually reaches a stripe state, $P_{\rm
    str}(L)$, as a function of $1/\log_2 L$ for $L$ up to 512.  Each data
  point, with error bars smaller than the size of the symbol, is based on
  $\geq 10^5$ initial spin configurations.  
\label{stripe}}
\end{figure}

Our simulations indicate that the system with zero initial magnetization
reaches a stripe state with a non-zero probability as $L\to\infty$
(Fig.~\ref{stripe}).  Linear extrapolation of the last 4 data points for the
probability of reaching the stripe state, $P_{\rm str}(L)$, versus $L$ gives
$P_{\rm str}(\infty)\approx 0.315$ and $0.344$ on the square and triangular
lattices, respectively.

The stripe state can contain an arbitrary even number of stripes.  For the
square system, we typically obtain two stripes of similar widths.  Metastable
states with more than two stripes appear very rarely.  For example, the
probability of reaching the four-stripe state grows very slowly with $L$ and
is less than $0.07\%$ for $L=200$.  We can gain a qualitative understanding
for the dependence of the probability of obtaining $k$ vertical stripes in
the final state, $P(k,A)$, by analyzing a general rectangular system of size
$AL\times L$ with fixed aspect ratio $A$ (and taking the $L\to\infty$ limit
as usual).  For example, for $A=9$ and $L=32$ the final probabilities
$P(k,A)$ are approximately 0.0028, 0.101, 0.35, 0.36, and 0.15, for $k=0,
2,\ldots, 8$.  There is also a probability approximately 0.034 that a
horizontal stripe forms.  In general, the probability $P(k,A)$ appears to be
peaked around $k_{\rm max}\propto A$.  Invoking the natural assumption of
analyticity in $A$ then implies that the probabilities $P(k,A)$ of $k$-stripe
states are positive for all (even) $k$ and arbitrary aspect ratio $A$.

Our data are insufficient to probe the $k$-dependence of $P(k,A)$ but
qualitatively $P(k,A)$ decays faster than exponentially.  This behavior seems
to be similar to that of the number of spanning clusters on the rectangles of
fixed shape at the percolation threshold.  In that problem, a wide consensus
that only one spanning cluster exists has been recently disproved by
numerical\cite{num} and theoretical\cite{theor} evidence.  We now employ an
argument similar to the one of Ref.~\cite{theor} in the context of spanning
clusters to estimate the large-$k$ behavior of $P(k,A)$.  Specifically, we
consider the probability $P_h(k,A)$ to reach a state with $k$ {\em
  horizontal} stripes (all $k$ stripes in the direction of the length $AL$).
Imagine now dividing the rectangle into two equal rectangles of size
$AL\times L/2$ each.  In the large-$k$ limit, the dominant contribution to
$P_h(k,A)$ comes from situations where approximately $k/2$ stripes traverse
each of the rectangles.  This implies
\begin{equation}
\label{est}
P_h(k,A)\sim [P_h(k/2,2A)]^2,
\end{equation}
which may be iterated to give
\begin{equation}
\label{esti}
P_h(k,A)\sim [P_h(2,Ak/2)]^{k/2}.
\end{equation}

The quantity on the right-hand side $P(2,Ak/2)$ is the probability to have a
stripe which traverses a rectangle of dimension $AL\times 2L/k$ in the long
direction.  Here we have assumed that the probabilities $P_h(k,A)$ depend
weakly on the system size $L$ and do not vanish in the thermodynamic limit.
If we take $L\sim k$ as the width of the original rectangle, then the
rectangle $AL\times 2L/k$ has a small finite width and a length of order
$Ak$.  This rectangle is so narrow that a stripe can occur only if existed in
the initial state.  This clearly occurs with probability $2^{-2\times {\rm
    length}}\sim e^{-Ak}$.  By substituting $P(2,Ak/2)\sim e^{-Ak}$ into
Eq.~(\ref{esti}) we deduce that
\begin{equation}
\label{estim}
P_h(k,A)\sim e^{-{\rm const.}\times Ak^2}.
\end{equation}
Thus $P(k,A)$ has a Gaussian tail. This explains why four and higher-stripe
states are almost never seen in our simulations for the square system.  In
the following, we always consider square (or hypercubic) systems.

\begin{figure}
  \narrowtext\hskip 0.3in\epsfxsize=2.2in\epsfysize=2.2in
\epsfbox{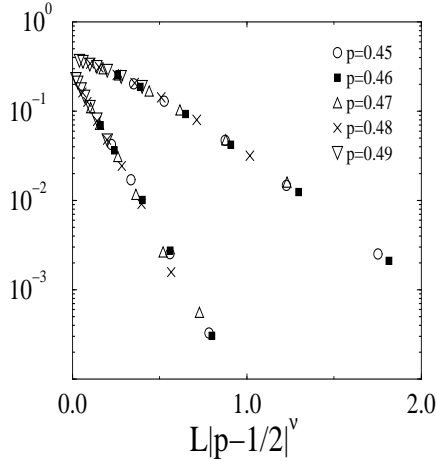}
\vskip
  0.25in
\caption{Probability that the minority phase wins (lower) or a stripe state
  is reached (upper) as a function of $L|p-1/2|^\nu$, with $\nu=1.5$ and
  $\nu=1.35$ respectively.  Data are based on systems with $L\leq 200$ and
  $\geq 5\times 10^4$ initial configurations.
\label{minwin2d}}
\end{figure}

Consider now an Ising system with a small difference between the number of up
and down spins.  We study two basic quantities: (i) the probability ${\cal
  M}(p,L)$ that the minority phase wins, that is, the sign of the
magnetization in the final (ground) state is opposite to that in the initial
state, and (ii) the probability ${\cal S}(p,L)$ that the system reaches a
stripe state.  Both these quantities exhibit scaling when $L$ diverges and
the the initial magnetization $m_0=2p-1$ vanishes such that the combination
$z\equiv L|2p-1|^{\nu}$ is kept constant.  For ${\cal M}(p,L)$, the best data
collapse is achieved with $\nu\approx 1.5$ while for ${\cal S}(p,L)$ the best
data collapse is achieved with $\nu\approx 1.35$.  Further, ${\cal M}(z)$
appears to be an exponentially decaying function of $z$ while ${\cal S}(z)$
appears to decay even more quickly (Fig.~\ref{minwin2d}).  The exponential
behavior is not surprising in view of the analogy to the spanning probability
in percolation.  Indeed, consider the extreme case case of percolation with
$p\to 0$.  Then a spanning cluster of the minority phase exists with
probability $\propto p^L$, {\it i.\ e.}, it decays exponentially with system
size.  This argument makes it plausible that ${\cal M}(z)$ and ${\cal S}(z)$
also decay at least as fast as exponentially with $L$ and thence with $z$.

Overall we conclude that in two dimensions, the initial majority phase always
wins in the limit $L\to\infty$.  Thus $P(p)=\theta(p-1/2)$ in two dimensions,
just as in the mean-field limit.  This suggests that $P(p)$ may well be a
step function for all spatial dimensions $d\geq 2$ when $L=\infty$.

\subsubsection{Finite magnetic field}

\begin{figure}
  \narrowtext \epsfysize=0.25in \hskip 0.8in\epsfbox{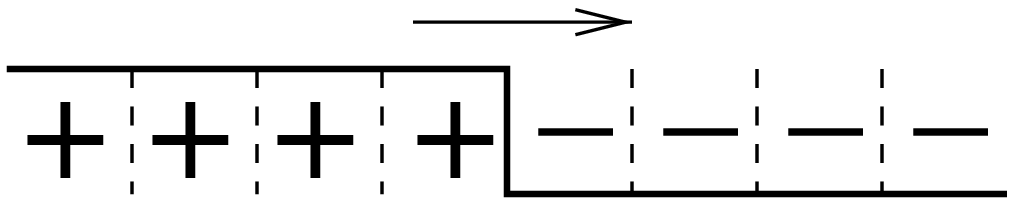} \vskip
  0.15in
\caption{In a weak positive field an interface kink can move only to the
  right, while in zero field this kink diffuses.
\label{kink}}
\end{figure}

It is also instructive to investigate the effect of a finite external
magnetic field $h$ on the fate of the system.  In two dimensions there are
two distinct field ranges: (i) weak fields $0<h<2J$, and (ii) strong fields
$2J<h$.  A weak field modifies the dynamics of spins which have two
misaligned neighbors.  In zero field such spins flip with rate 1/2, while in
a weak field they can only flip parallel to the field.  This means, for
example, that kinks on interfaces move in only one direction rather than
diffusing (Fig.~\ref{kink}).  For strong fields, down spins which have 3
misaligned neighbors can now flip parallel to the field and the system ends
up in the field-aligned ground state.

\begin{figure}
  \narrowtext \epsfxsize=1.6in \hskip 0.7in\epsfbox{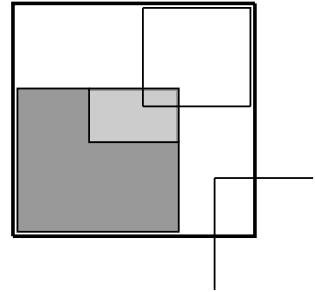} \vskip
  0.15in
\caption{Expansion of a cluster of up spins (dark shaded) in a weak
  magnetic field $h<2J$.  The convex hull (union of dark and light shaded
  regions) overlaps with the cluster on the upper right.  The convex hull of
  this aggregate (outer rectangle) then overlaps with yet another cluster,
  leading to continued expansion.
\label{expansion}}
\end{figure}

The most interesting case is that of weak field and a small initial
concentration of up spins, where the system consists of small clusters of up
spins in a background of down spins.  Due to the unidirectional kink motion
on interfaces, clusters of up spins can grow until each fills out its convex
hull (Fig.~\ref{expansion}).  If the convex hull of one cluster overlaps
with either another up cluster (or its convex hull), then the resulting
aggregate can expand further to fill out this enlarged convex hull.  If there
is yet another cluster (or convex hull) within this expansion zone, growth
continues.

This growth process is essentially the same as bootstrap percolation
\cite{bootstrap}, in which a lattice is randomly occupied, say with 
initial density $\rho_0$, and then all sites that do not have at least $n$
occupied neighbors are removed.  This deletion step is then 
repeated until no more
sites can be removed.  The case $n=3$ is essentially identical to our
weak-field system, with spins antiparallel to the field playing the role of
occupied sites in bootstrap percolation.  In $n=3$ bootstrap percolation, all
occupied sites will eventually be removed for $L\to\infty$, even if $\rho_0$
is arbitrarily close to 1.  Translating this to the Ising system, we conclude
that for any non-zero concentration of up spins, the system will evolve to
the ground state with $m=1$ in the thermodynamic limit.

\subsection{Three Dimensions}

\subsubsection{``Blinkers'' in zero magnetic field}

\begin{figure}
  \narrowtext \epsfxsize=2.0in\hskip 0.5in\epsfbox{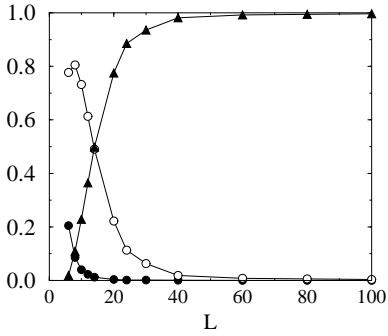} \vskip
  0.15in
\caption{Probability that the system reaches the ground state (dots), a frozen
  metastable state (circles), or a blinker state (triangles) as a function of
  the linear dimension.  The number of realization is $\geq 10^4$ for each
  system size.  The lines are a guide for the eye.
\label{final}}
\end{figure}

On the simple cubic lattice, the probability to reach the ground state,
$P_{\rm gs}(L)$, decreases rapidly with the system size (Fig.~\ref{final}).
For example, $P_{\rm gs}(L)\approx 0.001$ for $L=30$.  Perhaps even more
surprising is that the final state of the system for large $L$ is {\em not\/}
geometrically static!  Instead, the system contains blinkers
-- localized sets of spins which can flip indefinitely with no
energy cost.

To appreciate the nature of blinkers, we first describe the geometry of the
frozen metastable states.  For initial concentration of up and down spins
$p=1/2$ (which is much greater than the percolation threshold $p_c\approx
0.3116$ \cite{perc}), both the up and down spins percolate in all three
coordinate directions in the final state.  For example, for cubes with
$L=20$, 30 and 40, the probability that both phases percolate in all three
directions equals 0.83, 0.92 and 0.97 respectively.  Numerically the number
of distinct spin clusters almost always equals 2 -- there are no small
clusters of spins.

\begin{figure}
  \narrowtext \epsfxsize=1.3in\hskip 0.9in\epsfbox{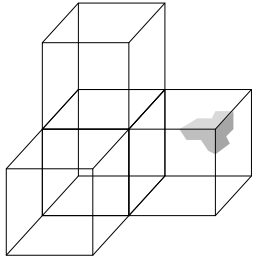} \vskip
  0.15in
\caption{A 3-arm star which percolates in all coordinate directions.  
  This cluster contains four cubic blocks of linear dimension $L/2$ in an
  $L\times L\times L$ system with periodic boundary conditions.  The other
  phase occupies the remaining half of the cube volume.  A convex corner (for
  free boundary conditions) is shown shaded.
\label{3-arm}}
\end{figure}

To visualize these percolating spin clusters, consider each spin as occupying
a unit cube.  A cluster must then have no convex corners to be stable
(Fig.~\ref{3-arm}); a spin at such a corner can flip freely and generate
additional convex corners.  The generic configuration which permits two
clusters of oppositely-oriented spins, each devoid of convex corners, to
percolate in all three directions has the form of two interpenetrating 3-arm
stars (Fig.~\ref{3-arm}).  Each arm is oriented along one coordinate
direction and joins onto itself because of the periodic boundary condition,
so that there are no convex corners.  This is the 3-dimensional analog
of stripes on the square lattice.

\begin{figure}
  \narrowtext \epsfxsize=1.8in\epsfysize=1.47in \hskip
  0.7in\epsfbox{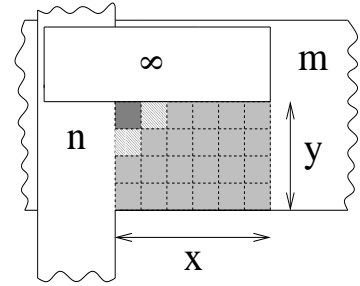} \vskip 0.15in
\caption{Stochastic blinker on the cubic lattice.  The shaded region of size
  $x\times y\times (n-m)$ contains a fluctuating interface which can range
  between all spins up and all spins down.
\label{blinkers}}
\end{figure}

{}From the 3-arm star, a blinker arises when the arms have different
cross-sectional areas, as sketched in Fig.~\ref{blinkers}.  Here we view the
percolating cluster of up spins as a building with an $m$-storey section
(marked $m$), an adjacent $n$-storey section (with $n>m$), and a section
(marked $\infty$) which wraps around the torus in the vertical direction and
rejoins the building on the ground floor.  The wiggly lines indicate wrapping
around the torus in the $x$- and $y$-directions.  This 3-arm star structure
has no convex corners and thus cannot shrink under Glauber kinetics.  The
shaded portion of Fig.~\ref{blinkers} supports a blinker.  This blinker
starts at the upper left corner of the shaded region of height $m$, where it
costs zero energy to flip the heavy shaded spin.  Once this spin flips, its
three nearest neighbors (two light shaded and one just above the dark-shaded
spin in Fig.~\ref{blinkers}) can also flip with no energy cost.  Continuing
this process gives rise to a fluctuating interface in the shaded
parallelepiped that is bounded by the states of all spins up and all spins
down.  Thus a blinking state wanders forever by transitions between connected
metastable configurations with the same energy.

\begin{figure}
  \narrowtext \epsfxsize=2.6in \hskip 0.2in\epsfbox{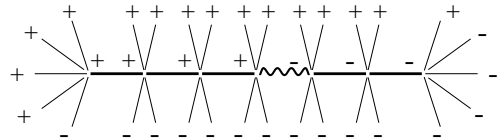} \vskip
  0.15in
\caption{Illustration of blinker topology on a 6-coordinated Cayley tree.  
  The blinker, namely, the bond between the misaligned spins (wiggly line)
  undergoes a random walk.
\label{tree-blink}}
\end{figure}

Blinkers can be visualized more simply on an even-coordinated Cayley tree
(Fig.~\ref{tree-blink}).  Consider, for example, a 6-coordinated tree in
which each spin in a finite segment (thick lines) feels zero exchange field
from the 4 neighboring spins which belong to separate chains.  The ends of
this segment are terminated by 5 additional chains with at least 4 spins in
agreement for these chains.  This segment is effectively a finite
one-dimensional chain with the endpoints fixed to be in opposite states so
that a blinker lives forever.

Amusingly, blinkers are much more prominent in the Potts model with Glauber
kinetics, where they arise even in two dimensions.  The Hamiltonian of the
$Q$-state Potts model is ${\cal H}=-J\sum_{\langle
ij\rangle}\delta_{\sigma_i\sigma_j}$.  Here $\sigma_i$ denotes the Potts
variable at site $i$ which can assume $Q$ distinct values and the sum is over
all nearest-neighbor pairs of sites\cite{potts}.  The zero-temperature
Glauber kinetics is analogous to the Ising case, {\it i.\ e.}, a spin flips
to agree with the majority of its neighbors.  In cases of a ``tie'', flips
occur with equal rates; for example, if half the neighbors of a given spin
are in one state and the other half are in a different state then the spin
flips to one of these two states with equal probability.

\begin{figure}
\vskip -0.05in
  \narrowtext \epsfxsize=1.3in\hskip 0.7in\epsfbox{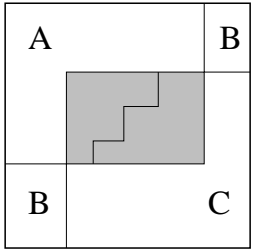} \vskip
  0.15in
\caption{Generic stochastic blinker in the kinetic three-state Potts model 
  on the square lattice.  The shaded region can flip between all spins in the
  $A$ and $C$ states, while all the remaining spins are stable since they
  have at most one misaligned neighbor.  A representative position of the
  $AC$ interface in the shaded region is shown.
\label{potts}}
\end{figure}

To appreciate the existence of blinkers, consider the three-state Potts model
in two dimensions and assume that all three phases have the same initial
concentration (this again corresponds to a $T=\infty$ initial state).  This
system can reach a metastable state which contains blinking domains, as
illustrated in Fig.~\ref{potts}.  Similar to the case of the Ising system on
the cubic lattice, the shaded region can blink between having all spins in
the $A$ or the $C$ states.  While it was previously argued\cite{lif,saf} that
domains can become pinned for quenches to $T=0$ when $Q\geq 3$, our
simulations of the $Q=3$ Potts model indicate that the probabilities to reach
the ground state and a frozen state decrease with the system size and
apparently approach to zero as $L\to\infty$.  In short, the Potts system gets
pinned in a blinking state rather than in a frozen state.

\subsubsection{Finite magnetic field}

\begin{figure}
  \narrowtext \epsfxsize=1.5in \hskip 0.7in\epsfbox{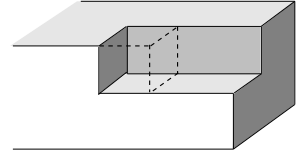} \vskip
  0.15in
\caption{Cluster of up spins with a concave trough.  This is sequentially 
  filled in, for $0<h<2J$, by flipping spins which each have 3 misaligned
  neighbors.  The dashed cube represent the spin which is about to flip.
\label{3d-convex}}
\end{figure}

An external magnetic field again drastically changes the final state of the
system.  On the cubic lattice there are now two critical fields, $h_1=2J$ and
$h_2=4J$, which demarcate different behaviors.  The regime $0<h<2J$
corresponds to $n=4$ bootstrap percolation for the clusters of spins
antiparallel to the field.  In the language of the Glauber kinetics, this
means that spins with initially three misaligned neighbors must align with
the field.  These spins, when flipped, fill in concavities and eventually
complete convex corners, as illustrated in Fig.~\ref{3d-convex}.

For $0<h<2J$, there is a threshold initial concentration of up spins, $p_h$,
below which finite droplets of up spins, with each spin having at least three
aligned neighbors, freeze.  For $p>p_h$, up spins eventually percolate due to
the infilling of concavities, which leads to the merging of clusters of up
spins and ultimately the ground state is reached.  Our simulations give $p_h$
significantly smaller than $p_c\approx 0.3116$ for site percolation on the
cubic lattice, in qualitative agreement with Ref.~\cite{bootstrap}.

\begin{figure}
  \narrowtext \hskip 0.7in\epsfxsize=1.5in\epsfbox{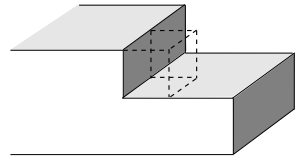} \vskip
  0.15in
\caption{Cluster of up spins with a concave interface.
  The energy cost of flipping the indicated spin is $4J-2h$; thus, 
  this spin will flip when $h>2J$.
\label{3d-concave}}
\end{figure}

The regime $2J<h<4J$ corresponds to $n=5$ bootstrap percolation.  It is now
possible to flip a spin adjacent to a straight but concave interface
(Fig.~\ref{3d-concave}).  This filling ultimately allows a cluster of up
spins to systematically expand and fill its convex hull.  This then leads to
a similar picture to $n=3$ bootstrap on the square lattice, in which
coalescence of convex hulls of neighboring clusters leads to a final state
where all spins are aligned with the field for any initial magnetization.
Finally for $h>4J$, even a single up spin nucleates the growth of additional
up spins and the system quickly reaches the ground state with all spins
pointing up.

\subsection{Final Magnetization Distribution}

In two dimensions, we have seen that the final state is either the ground
state or a stripe state in which there are typically two stripes of
approximately the same width.  This qualitative observation can be made more
precise by studying the magnetization distribution of the final state.
Systems which reach the ground state give $m=+1$ or $-1$, while systems which
reach the stripe state lead to a continuous component of the final
magnetization distribution which is peaked about 0.  The width of this peak
gradually narrows as the system size increases, but appears to converge to a
finite limit as $L\to\infty$.  

\begin{figure}
 \narrowtext\hskip -0.1in\epsfxsize=1.62in\epsfysize=1.7in\epsfbox{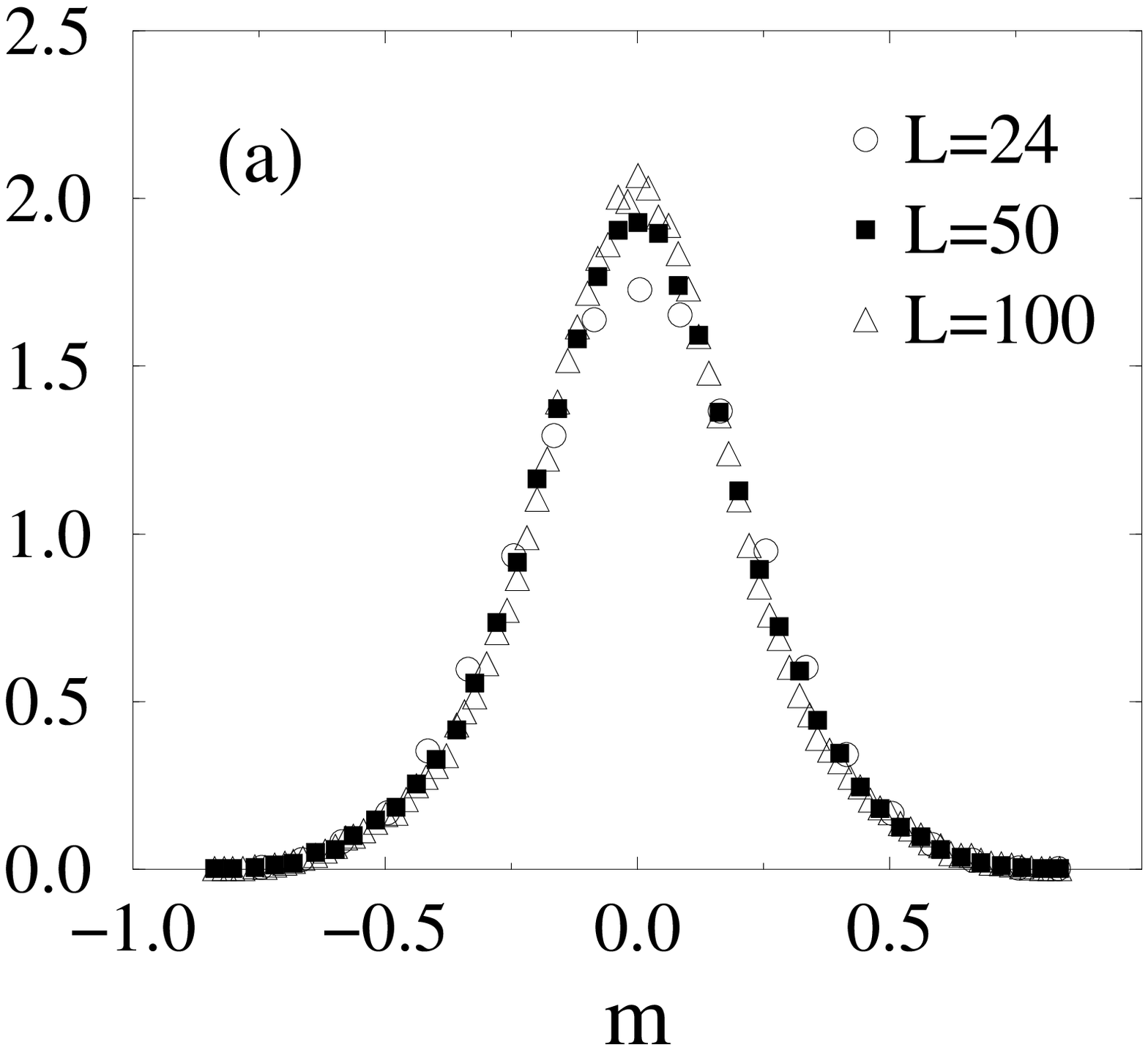}
 \hskip 0.03in\epsfxsize=1.62in\epsfysize=1.75in \epsfbox{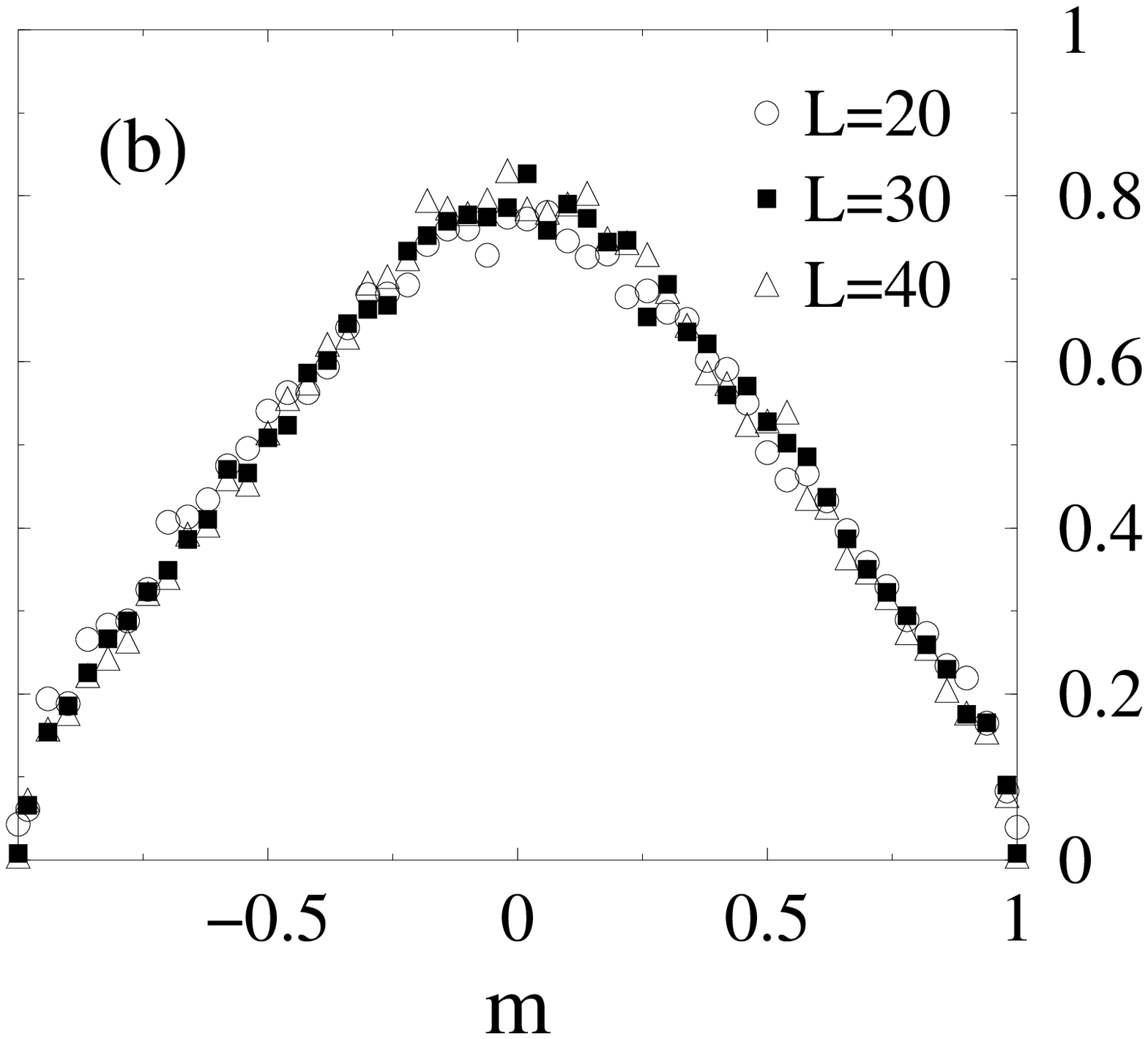}
\vskip
  0.25in
\caption{Final magnetization distribution in (a) $d=2$ and (b) $d=3$.
  The number of configurations is $\geq 10^5$ for every system size in $d=2$,
  and $\geq 5\times 10^4$ in $d=3$.  In $d=2$, the delta-function peaks
  at $m=\pm 1$ have been suppressed.
\label{mag}}
\end{figure}

In three and higher dimensions, the probability of reaching the ground state
is vanishingly small so that there is no longer delta-function peaks in the
final magnetization distribution at $m=\pm 1$.  In three dimensions, the
magnetization distribution has a relatively broad peak compared to two
dimensions.

\subsection{Final Energy Distribution}

\begin{figure}
  \narrowtext \epsfxsize=2.4in\hskip 0.3in\epsfbox{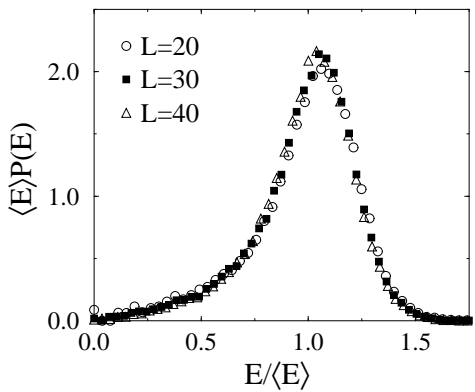} \vskip
  0.25in
\caption{Distribution of the normalized final energy per spin.  Number of
  configurations is $\geq 5\times 10^4$ for all system sizes.
\label{escale}}
\end{figure}

In two dimensions the distribution of the final energy of the system is a
series of delta-function peaks which correspond to configurations with
$0,~2,~4\ldots$ stripes.  In contrast, in three and higher dimensions the
energy distribution is continuously distributed and exhibits scaling in the
variable $E/\langle E\rangle$, where $E$ is the energy per spin (with the
ground state energy defined to be zero), and $\langle E\rangle$ its average
value.  (Fig.~\ref{escale}). 

For $d=3$, the average energy per spin $\langle E\rangle$ appears to decay as
$L^{-\chi}$, with the exponent $\chi\approx 1$.  Thus the total energy of the
final state, which is proportional to the total interfacial area, grows as
$L^2$.  This is consistent with the qualitative picture for the geometry of
the final state given in Fig.~\ref{3-arm}. 

\section{Number of Metastable States}

It is instructive to determine the number of metastable states as a function
of the spatial dimension because this helps quantify the relative influence
of these states on the evolution of the system.  The number of metastable
states can be computed asymptotically on the square lattice and on a
3-coordinated Cayley tree, with the latter providing an estimate for the
number of metastable states when $d=\infty$.  For $2<d<\infty$, we give a
simple lower bound for the number of metastable states which we expect gives
the correct asymptotic behavior.

\subsubsection{Two dimensions}

The metastable states of the ferromagnetic Ising-Glauber model on an $L\times
L$ square lattice with periodic boundary conditions consist of purely
vertical or horizontal stripe arrays, with the width of each stripe greater
than or equal to 2.  These states are essentially identical to the ground
states of the axial next-nearest neighbor (ANNNI) Ising chain with
nearest-neighbor ferromagnetic interaction $J_1$ and second-neighbor
antiferromagnetic interaction $J_2$ when $J_2=-J_1/2$.  For the ANNNI chain
with free boundaries, the number of metastable states was previously found in
terms of the Fibonacci numbers\cite{redner}.  For a chain of $L$ sites, the
number of these states grows asymptotically as $g^L$, where
$g=(1+\sqrt{5})/2$ is the golden ratio.

To determine the number of metastable states on an $L\times L$ square with
periodic boundary conditions, we need to account both for the fact that
stripes can be vertical or horizontal as well as the periodic boundary
conditions.  The former attribute means that the number of metastable states
on the square is twice that on a periodic one-dimensional chain.  The
periodic boundary condition also means that states which differ by overall
translation are not distinct.  This reduces the number of metastable states
of a periodic system by a factor of $1/L$ compared to free boundary
conditions.  Asymptotically, then, the number of metastable states on a
square of ${\cal N}=L^2$ sites is given by $M_2({\cal N})\sim
e^{B_2\sqrt{\cal N}}$, where $B_2=\ln g$.

\subsubsection{Dimensions $d>2$}

As already discussed, metastable states are geometrically more complex in
greater than two dimensions and their enumeration appears to be a difficult
problem.  However, we can give a simple lower bound for the number of
metastable states by constructing a higher-dimensional analog of the stripe
states.  In three dimensions, consider states which consist of an arbitrary
array of straight filaments such that each filament cross-section has size
$x\times y$ with $x,y\geq 2$, and that the distance between any two filaments
in either coordinate direction is also $\geq 2$.  The number of these
filamentary metastable states scales as $\exp(C_3L^2)$, where $C_3$ is a
constant.  Thus the existence of filamentary gives the lower bound for the
number of metastable states in three dimensions, $M_3({\cal N})>\exp(C_3{\cal
  N}^{2/3})$.  The analogous construction in $d$ dimensions yields $M_d({\cal
  N})>\exp(C_d{\cal N}^{1-1/d})$.

While we have not succeeded in constructing an upper bound for the number of
metastable states, it is plausible that this bound has the same form as the
lower bound.  In general, metastable states must consist of long filamentary
structures to avoid having any convex corners which serve as the nucleus for
energy-lowering moves.  This geometric constraint suggests that the
degeneracy of all metastable states should be similar to that of the lower
bound filamentary states.  Thus we expect that $M_d({\cal N})\sim
\exp\left(B_d {\cal N}^{1-1/d}\right)$.

\subsubsection{Infinite spatial dimension}

To probe an infinite spatial dimension, we estimate the number of metastable
states by considering the Cayley tree.  There is already a subtlety which
depends on whether the coordination number of the tree is even or odd.  While
odd-coordinated lattice systems exhibit the pathology of metastable freezing
for any initial magnetization, even-coordinated trees naturally give rise to
blinker states, as mentioned in the previous section.

\begin{figure}
  \narrowtext \epsfxsize=2.8in \hskip 0.1in\epsfbox{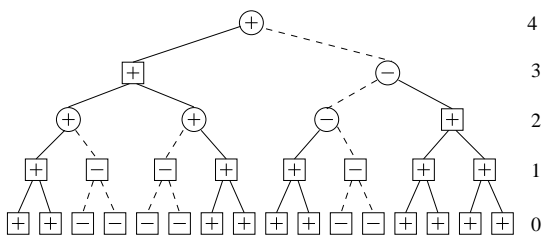} \vskip
  0.15in
\caption{A typical metastable state on a 3-coordinated Cayley tree.
  Circled spins are those whose state is determined by its parent one level
  higher while boxed spins are uniquely determined by the states of the
  daughter spins.  Clusters of negative spins are joined by dashed lines.
\label{tree}}
\end{figure}

For simplicity we consider the 3-coordinated tree with root at level $N$
(Fig.~\ref{tree}).  Two nodes in level $N-1$ are attached to the root site,
then four nodes form level $N-2$, {\it etc}.  To enumerate the total number
of metastable states $M_N$ on an $N$-level tree we first note that there are
two types of spins in any metastable state.  For a spin at level $n$, if the
two daughter spins in level $n-1$ agree, then the state of the parent spin is
uniquely determined.  Conversely, if the daughters disagree, then the state
of the spin in level $n$ (circled in Fig.~\ref{tree}) is determined by that
of its parent in level $n+1$.  (The neighboring spins at level 0 must agree.)

Let $D_N$ and $U_N$ are the respective number of metastable states with the
root spin determined and undetermined by its daughters.  Then by enumerating
all possible daughter states and the outcome of the root site, $D_N$ and
$U_N$ obey the recursion relations
\begin{eqnarray}
\label{reca}
D_{N+1}&=&{1\over 2}D_{N}^2 +{1\over 2}U_{N}^2 + 2D_{N}U_{N},\\
\label{recb}
U_{N+1} &=& D_{N}^2,
\end{eqnarray}
subject to the initial conditions $D_0=0$ and $U_0=2$.  For example, the
recurrence (\ref{recb}) expresses the fact that the root spin is undetermined
only when two daughter spins are uniquely determined and of opposite sign.
Thus $U_{N+1}=2\times D_N\times\frac{1}{2}D_N$, where the factor $2$ takes
into account that the root spin remains undetermined and the factor
$\frac{1}{2}$ ensures that the root spins in the daughter trees are opposite.

By computing the first few terms in these recursion formulae, we see that
$U_N$ and $D_N$ grow very rapidly with $N$.  To probe the asymptotic behavior
of $U_N$ and $D_N$, we divide Eq.~(\ref{reca}) by $U_{N+1}$ and use
Eq.~(\ref{recb}) to find the following recursion for $\Lambda_N=D_N/U_N$:
\begin{equation}
\Lambda_{N+1}=\frac{1}{2}+\frac{2}{\Lambda_N}+\frac{1}{2\Lambda_N^2}
\equiv R(\Lambda_N).
\label{lambdarecur}
\end{equation}
The recurrence (\ref{lambdarecur}) for $\Lambda_N$ iterates the rational
function $R(\Lambda)$ whose fixed points, $\Lambda=R(\Lambda)$, are $-1$ and
$(3\mp\sqrt{17})/4$.  Only the positive fixed point $\Lambda=(3+\sqrt{17})/4$
is physically acceptable.  This is also an attractive fixed point as
$R'(\Lambda)= -2\Lambda^{-2}-\Lambda^{-3}=(19-5\sqrt{17})/2=
-0.807764\ldots$, {\it i.\ e.}, $|R'(\Lambda)|<1$. Thus,
$\Lambda_N\to\Lambda$ and hence
\begin{equation}
\frac{D_N}{U_N}\rightarrow \Lambda\equiv \frac{3+\sqrt{17}}{4}.
\label{ratio}
\end{equation}
To determine $D_N$ we iterate $D_N=D_{N-1}^2 \Lambda_N$ to give
\begin{equation}
D_N=(D_1)^{2^{N-1}}\prod_{k=2}^{N}(\Lambda_{k})^{2^{N-k}}.
\label{DN}
\end{equation}
This equation implies that the following limit 
\begin{equation}
\label{defdelta}
\delta=\lim_{N\to\infty} (D_N)^{2^{-N}}
\end{equation}
exists and equals 
\begin{equation}
\label{delta}
\delta=\sqrt{D_1}\prod_{j=2}^\infty (\Lambda_j)^{2^{-j}}.
\end{equation}
{}From Eqs.~(\ref{reca})--(\ref{recb}) with initial conditions $D_0=0$ and
$U_0=2$, we obtain $D_1=2$ and $\Lambda_2=1/2$.  Using these values and the
recurrence (\ref{lambdarecur}) we numerically determine $\delta=
1.56581199\ldots$.  

Equation (\ref{defdelta}) gives $D_N\propto \delta^{2^N}$ but
we can also find the overall amplitude.  {}From Eqs.~(\ref{DN}) and
(\ref{delta}) we find the exact expression for the ratio
\begin{equation}
\label{rat}
\delta^{2^N}/D_N=\prod_{j=1}^\infty (\Lambda_{j+N})^{2^{-j}}.
\end{equation}
Now we recall that $\Lambda_N\to \Lambda$ and thus the product on the
right-hand side of Eq.~(\ref{rat}) approaches to $\Lambda$ as
$N\to\infty$. This together with Eq.~(\ref{ratio}) yields 
\begin{equation}
\label{DUN}
D_N\to \Lambda^{-1}\delta^{2^N}, \quad 
U_N\to \delta^{2^N}.
\end{equation}
These results should be compared to the total number of the spin states
$S_N=2^{\cal{N}}$, where ${\cal N}=2^{N+1}-1$ is the total number of sites in
the $N$-level tree of coordination number 3.  Note also that the ``entropy''
of the total number of metastable states, $\ln (U_n+D_N)$, asymptotically
grows as $C{\cal N}$, with $C={1\over 2}\,\ln\delta\cong 0.224202$.  The
linear ${\cal N}$-dependence fits with the previous lower bound according to
which the metastable state entropy increases as ${\cal N}^{1-1/d}$ in $d$
dimensions.

\section{Finite temperature}

For a system which is quenched from infinite to a low but non-zero
temperature, the equilibrium state is eventually reached.  However, we find
that the approach to equilibrium proceeds in two distinct stages.  In the
initial coarsening stage, the evolution is essentially the same as that of
the zero temperature case.  In two dimensions, for example, the system first
relaxes to a metastable stripe state with probability $\approx 1/3$.  At zero
temperature, this would be the final state of the system.  However, at finite
temperature, there is a relatively slow escape from this metastable state
whose rate we now determine by a simple geometric approach.

\begin{figure}
  \narrowtext \hskip -0.1in\epsfxsize=3.3in\hskip
  0.0in\epsfbox{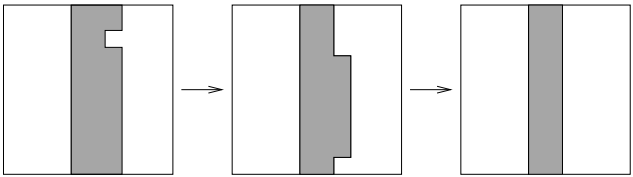} \vskip 0.15in
\caption{Relaxation of a stripe state in two dimensions at small non-zero
  temperature: (a) Nucleation of a dent; (b) Diffusive growth of the dent;
  (c) Dent reaches the system size and hence the domain wall steps
  to the left.  This overall process ultimately leads to the disappearance of
  the stripe.
\label{dent}}
\end{figure}

A stripe is formed in a time of order $L^2$\cite{skr}.  At a small positive
temperature, a stripe can disappear by the annihilation of the two domain
walls (Fig.~\ref{dent}).  This annihilation occurs by the following steps:
First, a dent is created by flipping a spin at a domain wall.  The time
required for this event is of the order of $e^{4J/kT}$, where $4J$ is the
energy cost associated with the spin flip.  Once a dent is created, the spin
in the dent, as well its two vertical neighbors, are now free to flip.  Thus
the length of the dent performs a one-dimensional random walk until the
horizontal boundaries meet.

Now using elementary facts about the first passage of a one-dimensional
random walk in the presence of an absorbing boundary\cite{fpp}, the dent
recombines with probability $(L-1)/L$ and the domain wall returns to its
original state, or with probability $1/L$ the dent expands and changes the
sign of one column of spins.  Thus we need $L$ dent creation events before
the interface of the strip hops rigidly by $\pm 1$ in the $x$-direction.  The
time needed for this hop is therefore of the order of $L\,\exp(4J/kT)$.
Since the typical width of a stripe is of the order of $L$, there must
typically be $L^2$ such interface hopping events before the two interfaces
meet and thus surmount the metastable barrier.  As a result, the time for a
stripe state to disappear is of the order of $L^3\,\exp(4J/kT)$.

\begin{figure}
  \narrowtext \hskip 0.4in\epsfxsize=2.0in\epsfbox{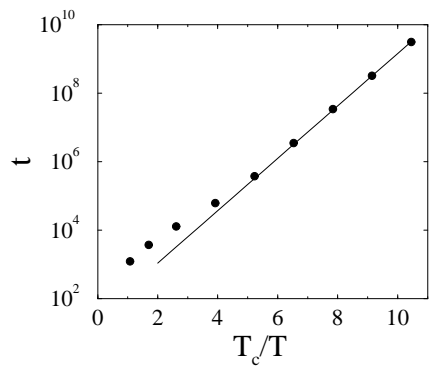}
\vskip  0.25in
\caption{Time to reach the equilibrium state on the square lattice.  The
  straight line is $L^3\, e^{4J/kT}$.
\label{nonzerot}}
\end{figure}

Our simulations are in excellent agreement with this prediction for
$T/T_c\alt 0.2$ (Fig.~\ref{nonzerot}).  Here the time to reach the
equilibrium is defined as the time for the system to first reach the
equilibrium value of the magnetization of the Ising model on the square
lattice at temperature $T$.  This stopping time is dominated by
configurations which first reach a stripe state; configurations which relax
directly to the equilibrium state reach this state quickly.

We can develop a similar argument in three dimensions.  Since a typical
metastable state has the form of two interpenetrating 3-arm stars, let us
estimate the time for such a structure to disappear.  The lowest energy
excitation is to flip a spin in one of the concave corners of this structure
(see Fig.~\ref{3d-concave}).  This spin flip requires an energy of $4J$ and
thus takes a time of the order of $\exp(4J/kT)$.  By flipping of the order of
$L^2$ such spins it is possible to create a planar barrier of one phase which
spans the system, at which point the other phase can disappear with no
additional energy cost.  Since the energy cost associated with the creation
of this excitation is $4JL^2$, this suggests that the time to surmount the
metastable barrier scales as $\exp(4JL^2/kT)$, a time which is too long to
probe directly by simulations.

\section{Summary}

We have investigated the evolution of a finite Ising system with Glauber
kinetics when it is suddenly quenched from infinite to zero temperature.  In
two dimensions there appears to be a non-zero probability that the system
ultimately reaches a frozen metastable state which consists of two or more
parallel straight stripes.  While our simulations suggest that the
probability of reaching a stripe state is positive even as $L\to\infty$, and
we have a heuristic argument that $k$-stripe states occur with positive
probability for every even $k$, we do not have a rigorous argument to support
this observation.  This is a fundamental unanswered question.

In three and higher spatial dimensions, the probability that system reaches
either the ground state or a frozen state is vanishingly small.  Essentially
all realizations end up wandering forever on connected iso-energy sets of
blinker states.  It is easy to visualize these blinkers on a even-coordinated
Cayley tree as well as on a small-size cubic lattice.  However, we do not
have a good way to characterize these blinkers for large finite-dimensional
systems.  The existence of blinkers means that the kinetic Ising-Glauber
system in sufficiently large spatial dimensions belongs to type ${\cal M}$
according to the classification of Newman and Stein\cite{ns}.  That is, some
fraction of the spins flip infinitely often (those on blinkers), while the
rest of the spins flip a finite number of times.

One reason for the system not reaching the ground state is that
metastable states become more numerous as the spatial dimension increases.
The number of these metastable states appears to scale as $\exp({\cal
  N}^{1-1/d})$ in $d$ dimensions, where ${\cal N}$ is the total number of
spins.  This makes it plausible that a spin system is more likely to first
encounter and get trapped in a metastable state before the ground state can
be reached.  Associated with the metastable states are a variety of
interesting geometric features of the final state, such as the distribution
of magnetization and the distribution of energy.  Many features of these
distributions are still unexplained.

At low temperature, the Ising-Glauber system necessarily reaches equilibrium,
but via a two-stage relaxation process.  Initially, the kinetics is nearly
identical to that of the zero-temperature case.  For the subset of systems
which reach a metastable stripe state in two dimensions, there is then a slow
approach to equilibrium by the nucleation of defects which cause the stripe
boundaries to diffuse, ultimately merge, and thus disappear.  Because this
kinetics is an activated process, the time to relax to equilibrium is
extremely slow.  Surprisingly, this two-stage picture persists for
temperatures up to approximately $0.2T_c$ in two dimensions.  A similar
two-stage picture appears to hold in higher dimensions.  However, the time
scales associated with surmounting metastable barriers and ultimately
reaching the ground state are astronomically long.

We are grateful to NSF grant No.~DMR9978902 for partial support of this work.

\end{multicols}

\begin{thebibliography}{}

\bibitem{glauber}
  R.~J.~Glauber, J. Math.\ Phys.\ {\bf 4}, 294 (1963).

\bibitem{rev}
  J.~D.~Gunton, M.~San~Miguel, and P.~S.~Sahni in: {\it
  Phase Transitions and Critical Phenomena}, Vol. 8, eds. C.~Domb and
  J.~L.~Lebowitz (Academic, NY 1983); A.~J.~Bray, Adv.\ Phys.\ {\bf 43},
  357 (1994).

\bibitem{ns}
  C.~M.~Newman and D.~L.~Stein, Phys.\ Rev.\ Lett.\ {\bf 82},
  3944 (1999); Physica A {\bf 279}, 159 (2000).

\bibitem{skr} A preliminary account of this work is given in V.~Spirin, 
P.~L.~Krapivsky, and S.~Redner, Phys.\ Rev.\ E {\bf 63}, 036118 (2001).

\bibitem{liggett}  See {\it e.\ g.}, T. M. Liggett, {\it Interacting Particle
  Systems} (Springer-Verlag, New York, 1985).

\bibitem{bootstrap}
  J.~Chalupa, P.~L.~Leath, and G.~Reich, J.\ Phys.\ C {\bf 12}, L31 (1979);
  P. M. Kogut and P. L. Leath, J.\ Phys.\ C {\bf 14}, 3187 (1981);
  J.~Adler, Physica A {\bf 171}, 453 (1991). 
  
\bibitem{perc} D. Stauffer and A. Aharony, {\it Introduction to
  Percolation Theory} (Taylor \& Francis, London, 1992). 
  
\bibitem{other} Anomalies associated with stripes and other metastable states
  in the Ising model have been hinted at in E.~T.~Gawlinski, M.~Grant,
  J.~D.~Gunton, and K.~Kaski, Phys.\ Rev.\ B {\bf 31}, 281, (1985);
  J.~Vi\~nals and M.~Grant, Phys.\ Rev.\ B {\bf 36}, 7036 (1987); J.~Kurchan
  and L.~Laloux, J. Phys.\ A {\bf 29}, 1929 (1996); D.~S.~Fisher, Physica D
  {\bf 102}, 204 (1997); A.~Lipowski, Physica A {\bf 268}, 6 (1999).  For the
  Potts model see \cite{lif,saf}.

\bibitem{lif} I.~M.~Lifshitz, Zh.\ Exp.\ Theor.\ Fiz. {\bf 42}, 
  1354 (1962) [Sov.\ Phys.--JETP {\bf 15}, 939 (1962)].

\bibitem{saf} S.~A.~Safran,
  P.~S.~Sahni, and G.~S.~Grest, Phys.\ Rev.\ B {\bf 28}, 2693 (1983);
  P.~S.~Sahni, D.~J.~Srolovitz, G.~S.~Grest, M.~P.~Anderson, and 
  S.~A.~Safran, Phys.\ Rev.\ B {\bf 28}, 2705 (1983).

\bibitem{num}
  P.~Sen, Int.\ J. Mod.\ Phys.\ C {\bf 8}, 229 (1997) and {\bf 10}, 747
  (1999); P.~Grassberger and W.~Nadler, {\it cond-mat}/0010265. 

\bibitem{theor}
  M.~Aizenman, Nucl.\ Phys. B {\bf 485}, 551 (1997);
  J.~Cardy, J. Phys.\ A {\bf 31}, L105 (1998).

\bibitem{potts}
  F. Y. Wu, Rev.\ Mod.\ Phys. {\bf 54}, 235 (1982).

\bibitem{redner}
  S.~Redner, J. Stat.\ Phys.\ {\bf 25}, 15 (1981).

\bibitem{fpp}
  S. Redner, {\it A Guide to First-Passage Processes} (Cambridge University
  Press, New York, 2001).

\end{thebibliography}
\end{document}